\documentstyle[12pt]{article}
        \def\be{\begin{equation}}
        \def\ee{\end{equation}}
        \def\v{\vert}
\begin{document}
\begin{center}
{\bf A Multi Species Asymmetric Exclusion Process,\\
Steady State and Correlation Functions on a Periodic Lattice}

\vspace {1cm}

{V. Karimipour} \\
\vspace {1cm}
School of Physics,\\
Sharif University of Technology\\
P.O.Box 11365-9161\\
and\\
Institute for Studies in Theoretical Physics and Mathematics\\
P.O.Box 19395-5746\\
Tehran, Iran\\
\vspace {1cm}

{Abstract} \\
\end{center}
By generalizing the algebra of operators of the Asymmetric
Simple Exclusion Process (ASEP),
a multi-species ASEP in which particles can overtake each other,is defined on both open and closed one dimensional chains.
On the ring the steady state
and the correlation functions are obtained exactly.
The relation to particle hopping models of traffic and the possibility
of shock waves in open systems is discussed. The effect of the boundary condition on the steady state properties of the bulk is studied.
\\ \\ \\
{\bf key words}: matrix product ansatz, operator algebra, asymmetric
 exclusion process,\\
traffic flow\\
{\bf Pacs numbers}:050.60.+w, 05.40.+j, 02.50.Ey
\newpage

\section{Introduction}
The aim of this letter is to introduce a natural multi-species generalization
of the one-species asymmetric simple exclusion process (1-ASEP) [1-5].
We will show that using the Matrix Product Ansatz (MPA) for one dimensional
stochastic systems, one can algebraicly define a natural p-ASEP ( p = number
of species) such that
all the results of 1-ASEP will be reproduced in the special
case $p=1$. We will also show that this model incorporates new features which
compared with 1-ASEP makes it a better candidate as a model of traffic flow.
First we resume the basic facts about 1-ASEP. In this process which is defined
on a one dimensional lattice with either periodic or open boundary conditions
, particles hop stochastically to their right nearest neighbor site if this
site is empty
and stop if this site is occupied.This model is related to many interesting
physical problems including interfacial growth [6], directed polymers in
random media [7,8],kinetics of biopolymerization [9], and
traffic flow [10-12] or any queing problem.
It is an example of a system far from equilibrium for which many exact results
are known (for recent reviews see [5,13]).Some of the basic results are
as follows.\\ On a ring of N sites where there
are M particles interacting with
each other, the steady state is such that all configurations have equal
weights [14].
All the equal time correlation
functions are also easy to calculate [14], that is
$$ <\tau_i > = { M\over N}\hskip 1cm <\tau_i \tau_j > = { M
(M-1)\over N (N-1)}
\ \ \ \ \ \ \ {\rm etc.}$$
Here the stochastic variable $ \tau_k $ is 0 if site k is empty and 1
otherwise.
Therefore $<\tau_k>$ is the average density and $<\tau_k \tau_l>$ is the
density-density correlation.
The unequal time correlation functions are not known in general and only
certain quantities, like the diffusion constants have been calculated exactly
[15,16].\\
On an open chain where particles arrive at the left with rate $\alpha$ and
leave
the right with rate $\beta$ the steady state and the phase diagram [17-19] is
known exactly. Particularly when
$\alpha + \beta = 1 $, the steady state is defined by a Bernouli measure.
The average density
of particles is given by $ \alpha $  and the steady current is $ J =
\alpha (1-\alpha)$.
The measure being multiplicative, there is no correlation in this case.\\
When $ \alpha + \beta \ne 1 $, the steady state is again multiplicative [19
], but this time
matrices replace numbers. This is the content of the matrix product ansatz
(MPA)[20]. In this case the probability of a configuration $ (
\tau_1,\tau_2,...,\tau_N)$ is given by
\be  P(\tau_1,\tau_2,...,\tau_N) = {1\over Z_N}< W \v
D(\tau_1)D(\tau_2)...D(\tau_N)\v V >\ee
where the matrices $ D:= D(1) $ and $ E:=D(0)$ and the vectors $ < W\v $
and $ \v V > $
satisfy the following relations:
\be DE= D+E \hskip 1cm D\v V> = {1\over \beta}\v V> \hskip 1cm <W\v E = {1\over
\alpha} <W\v \ee
and $ Z_N$ is a normalization constant.
(For the dynamical matrix ansatz see [21].)
It can be shown that the representations of the algebra (2) are either
one dimensional or infinite dimensional. The one dimensional representations
are the ones which give the steady state on the ring( where the matrix element
(1) is replaced by a trace),and on the open chain
when $ \alpha + \beta = 1 $. The infinite dimensional representations are used
for the open chain when
$ \alpha + \beta \ne 1 $. Although the details of the
one dimensional representations
are important for the properties of the steady state in the open chain,
for
the periodic chain where the number of particles are fixed, once the existence of a one dimensional representation is
shown, it follows that all configurations are of equal weights and calculation
of all equal time quantities can be done simply by using combinatorics.
In [22] it has been shown that the MPA is really not an ansatz and in fact the
steady state of any one-dimensional stochastic process governed by a Hamiltonian
with only nearest neighbor interaction can be put in the form (1). In the
general case
where each variable $ \tau_k $ takes $ p+1 $ values say $ 0, 1, ... p $, one
writes again the steady state, in the same form as in (1).
The p+1 operators $ D_i:= D(\tau_i)$  and the vectors $ \vert V> $ and $ < W\vert $ should satisfy a set of algebraic
relations [22] written compactly as follows:
\be h^B({\cal D}\otimes {\cal D}) = X\otimes {\cal D} - {\cal D}\otimes X \ee
\be (h^N{\cal D} - X) \vert V> = 0  \hskip 2cm
 <W\vert (h^1{\cal D} + X) =0\ee
where $ h^B $ is the part of the bulk Hamiltonian acting on two neighboring
sites, $h^1$ and $ h^N$ are boundary terms, and  $ ({\cal D})_i=D_i $ and $ X $ are two column matrices. Usually $
X $ is a suitably chosen numerical matrix.
However, finding the representations of the obtained algebra is by no
means an easy problem and in fact may well be harder than the original problem.
In this sence the work of Kreb and Sandow [22] implies that the MPA does not
facilitates much
the search for the steady state of one-dimensional stochastic systems or
the ground
state energies of generalized spin chains.However one can reverse the problem
and turns the observation of [22] into a virtue. Thats, one can postulate a
consistent
algebra with nontrivial representations and then find via (3-4),if a
physically meaningful process or spin Hamiltonian corresponds to this algebra.
In this letter we will go through this procedure and define the p-species
ASEP.
We start from
the following algebra
\begin{eqnarray} D_i E &=& {1\over v_i} D_i + E  \hskip 1cm i=1...p\\
 D_j D_i &=& {v_i D_j - v_j D_i \over {v_i - v_j }}\hskip 1cm j>i\end{eqnarray}
where the parameteres $ v_i$ are finite non-zero real numbers which we order as
$ v_1 \leq v_2 \leq .... \leq v_p $ .
It is easy to check that this algebra is associative, thats for any
$ k > j > i$,
$ D_j(D_i E) = (D_j D_i) E $ and $
 D_k(D_j D_i) = (D_k D_j) D_i $.
For details of the derivation and also for the representations see [23].
It is now straightforward to check that the following Hamiltonian when inserted
in (3) gives the above algebra,
\be h^B = -\sum _{i=1}^p v_i ( E_{0i}\otimes E_{i0} - E_{ii}\otimes E_{00})
- \sum _{j>i}^p (v_j - v_i ) (E_{ij}\otimes E_{ji} - E_{jj}\otimes E_{ii})\ee
where the choice made for $X$ is as
follows: \  $ x_0 = -1 , x_i = v_i/p $,
and ${\cal D}_0 = E , ({\cal D})_i = {D_i\over p} $.The factors $ {1\over p}
$ are for later convenience. Here the matrices $ E_{ij}$ have the standard definition
$ (E_{ij})_{kl} = \delta_{i,k}\delta_{j,l}$.
The local Hilbert space of each site is spanned by the
vectors $ \v 0 > , \v 1 >,... \v p > $. The state $\v 0 > $ means that the
corresponding site is vacant and the state $ \v j > $ means that it is occupied
by a particle of type $ j $.
The Hamiltonian (7) describes a process in which particles of type $ j $
hop with rate $ v_j $ and when they encounter particles of type $i $,
with $ v_i < v_j $
they interchange their sites, as if fast particles stochastically overtake slow
particles
with a rate $ v_j - v_i $. This model seems to be very natural as a simple
model of one-way traffic flow.
Real models of traffic flow are too complex to be amenable to analytical
treatment. One species ASEP is a very simple model of traffic which has been
extensively
studied in the past few years by analytical methods. Compared with 1-ASEP,  our
model,while still based on a nice algebraic structure and hence amenable to
analytical treatment, incorporates
new more realistic features of traffic flow, namely the existence of
different intrinsic speeds for the cars and the possibility of overtaking or
passing [24].
Moreover, while 1-ASEP is a suitable model for one-lane traffic flow
the present model, at least in those situations where the probability of
cars riding side by side is small, seems to be a good model for multi-lane
traffic flow.
Note that when fast particles with speed of say $
v_2$  reach slow ones with speed $ v_1 < v_2 $ they pass them only
 stochastically.
That is,in a time interval $ dt$ a fraction $ (v_2-v_1)dt$ of the fast cars
overtake  and the rest are stopped behind the slow ones. This means that
although a fast car has an intrinsic speed $ v_2 > v_1 $ in an empty road,its
effective speed depends on both the distance and the relative speed of the
car ahead. Considering a two particle system ( a fast one with
coordinate  $ x_2 $ behind a slow one with coordinat $ x_1$) we can write :
\be {v_2}^{eff}:= {d\over dt}< x_2> = v_2 - v_1 P(-1,t)\ee
where $ P(-1,t)$ is the probability of the fast particle being
one lattice site behind the slow particle. The above equation can be obtained
after lenghty calculations starting from the master equation of the two particle
system in coordinate space, although
it is obvious from the very definiton of the process.
It is intuitively clear that $ P(-1,t)$ is an increasing function
of $ v_2 - v_1 $ and a decreasing function of the initial seperation
$ x_1 - x_2 $.
which implies a realistic behaviour for $ {v_2}^{eff}$.\\
Moreover we have shown elsewhere [23] that the same algebra (5,6) describes
p-ASEP
on an open system where particles of speed $ v_i $ enter the left with rate
$ {\alpha v_i\over p}$
and leave the right with rate $ v_i +\beta -1$.
In the open system the unit of
time is set so that the average speed of particles is unity. Hence ${\alpha}$
and $ \beta $ are the total arrival and average departure rates respectively.
The forms of these rates and their dependence on speed again is a natural
property
of highway traffic flow.
So much for the relevance of this model to traffic flow.
\section{Correlation Functions}

In the following we
restrict ourselves to the periodic boundary condition and calculate the steady
state and some equal-time correlation functions.\\
The algebra (5,6) has a one parameter family of one dimensional representations,
namely $ E= {1\over \alpha}, \ \ D_i = {
v_i \over { v_i - \alpha }}$ where $ \alpha $ is a free parameter. This means
that in the steady state
all configurations have
equal weights. If there are $ m_i $ particles of type $i $, and the
total number of particles is $ M = m_1 + m_2 + ... m_p $,then the probability
of all the configurations
are equal to :
\be C = { 1\over N!} m_1! m_2!... m_p ! ( N- M)!\ee
We define the following functions:
\begin{eqnarray} n^{(i)}(\tau_k ) &:=& \delta_{i,{\tau_k}}\hskip 1cm \cr
n(\tau_k) &:=&n^{(1)}(\tau_k )+ n^{(2)}(\tau_k )+ ... n^{(p)}(\tau_k )  \cr
v(\tau_k) &:=&v_1 n^{(1)}(\tau_k )+ v_2 n^{(2)}(\tau_k )+ ... v_pn^{(p)}
(\tau_k )\end{eqnarray}
The one point functions $ < n^{(i)}_k>  , < n_k>  $ and $< v_k> $ of
these quantities give respectively the average number density of particles
of type (i),the average number density of
all types of particles and the average intrinsic speed of particles at site k.
Due to the uniformity of the measure none of the correlation functions
will depend
on the site indices.
It is simple to show the following:
\be < n^{(i)}_q > = {m_i\over N},\hskip 1cm
< n^{(i)}_q n^{(j)}_r >  = \cases{ {m_i(m_i -1)\over N(N-1)}
\hskip 1cm  i=j \cr {m_im_j \over N(N-1)} \hskip 1cm \ i\ne j\cr} \ee
Higher correlation functions have similar forms (i.e
$ < n^{(i)}_q n^{(i)}_r n^{(i)}_s >=  {m_i(m_i -1)(m_i - 2)
\over N(N-1)(N-2)}$ ).
As a sample calculation we note that:
\be < n^{(i)}_q> \equiv < n^{(i)}_1> = \sum_{ \{ \tau_{\alpha} \} }
\delta_{\tau_1,i}P(\tau_1,...,\tau_N)
=  C\sum_{\tau_2,\tau_3,...\tau_N}1 \ee
The last sum is the number of ways the rest of particles (except (i))
can be distributed on the ring. Combining this
with the value of C given in (9) gives the result.
From (11) the following quantities are calculated:
\be < n_q > = {M\over N} \hskip 1cm
 < v_q > = {M\over N}<v> \hskip 1cm
 < n_q n_r > = {M(M-1)\over N(N-1)}\ee
\be < v_q v_r > = { M^2 <v>^2 - M<v^2> \over N(N-1)}\ee
Here the averages $ <v>, and <v^2> $ are taken with respect to the
population present in the
system, i.e:
$$ < v > := {m_1 v_1+ m_2 v_2 + ... m_p v_p \over M}$$
while the left hand side averages are taken with respect to the steady
state configurations
of the particles on the ring.
The result (14) is the new quantity  which can be
defined in this process and is absent in 1-ASEP.
From (13) and (14) one obtains the correlation function for intrinsic speeds
( or in fact types of particles ) at different sites:
\be g(\rho):= <v_k v_l> -<v_k><v_l>= {\rho \over {N-1}}(\rho <v>^2 - <v^2>)\ee
where $ \rho = { M\over N} $ is the density .
It is seen that the maximum correlation exists at $ \rho = 0 $ and $ \rho =
{<v^2>\over <v>^2} $ and the minimum correlation at $ \rho = {<v^2>\over {2<v>^2}} $
All the other correlation functions can be calculated in this manner.For
example some lenghty calculation will give
\be < v_q v_r v_s  > = { M^3 <v>^3 - 3M^2<v><v^2> +2M<v^3> \over N(N-1)(N-2)}\ee

\section{Boundary Induced Negative Currents}

An intersting physical quantity to consider is the average current of each type of particles.
From the form of the process one can write the following continuity equation for the
density of each type of particles.
$$ {d\over dt}< n^{(i)}_k > = < J^{(i)}_k >-<J^{(i)}_{k+1}>$$
where the current of particles of type (i) is given by:
\be <J^{(i)}_k> = v_i <n^{(i)}_{k-1} \epsilon_{k}> + \sum_{j<i} (v_i - v_j)
<n^{(i)}_{k-1} n^{(j)}_{k} >
 -\sum_{j>i} (v_j - v_i) <n^{(j)}_{k-1} n^{(i)}_{k} > \ee
and
$ \epsilon_k = 1 - \sum_i n^{(i)}_k $. In fact  $ <n^{(i)}_{k-1}
 \epsilon_k >$ is the probability of site k-1 being filled with an (i)
particle and site k being empty. .
From (14,17) the average  current is calculated to be:
\be < J^{(i)}> = { m_i\over N(N-1)}\bigg( Nv_i - M <v>\bigg) \ee
Finally we obtain the total current as :
\be <J> := <J^{(1)} + J^{(2)} + ... J^{(p)} > = {M(N-M)\over N(N-1)}<v> \ee
again in accord with the 1-ASEP result.\\
The\  \ \ interesting\  point\  is\  that\  for\  those\  particles\  whose\
hopping\ \ \ \ \
 rates \ \ \ are
less \ \ \ than ${M\over N}<v>$, the currents $< J^{(i)}>$ turns out to be negative.
One may think that this result is expectable, once
we remind of the exchange of particles which takes place in the process i.e;
particles ordinarily hop forward but when encountered by faster particles from
the left they hop backward. However negative currents are induced only on the
ring and it has been proved rigourously [23] that in an open system
all the currents
are proportioanl to one single current. This is an example of how in a
nonequilibrium situation boundary conditions drastically affect the behaviour
 in the bulk.
A simple way for controlling negative currents by varying the density is as
follows: As far as
$ M < M_c:={N v_{min}\over <v>}$ where $ v_{min} $ is the lowest speed
of the particles
there is no negative current in the system. One can now add particles of
speed $ <v>$ to
the system which increases only $ M $ without changing $<v>$ . For all
$ M > M_c$
negative currents will be developed in the system.\\
We would like to stress that the merits of this inherent partial
asymmetry are quite
remarkable. On the one hand, one is tempted to think that this model
may also have some
relevance as a two-way traffic model and on the other hand, the algebra that
describes this process is much simpler than
the commutator-like algebras which one may try to define for a
multi-species partialy
asymmetric process from the begining. In fact the attempted algebras of
this later type [25,26] do not
have even a general and elegant form for arbitrary p.
\section{A note on Shock Waves}

For the following discussion we consider open boundary conditions. In this case
the one dimensional representations give the steady state for the case
 $ \alpha + \beta = 1$ [23].
The steady state is a translationaly invariant Bernouli measure in
which the density
of particles is constant along the chain. In the $ p\longrightarrow
\infty$ limit, where
the speeds of the particles are taken from a distribution $ P(v)$, the
total density
and the total currents are given by [23]:
\be \rho(\alpha)= {\Delta(\alpha) \over {{1\over \alpha} + \Delta(\alpha)}}
\hskip 1cm J(\alpha) =
{1 \over {{1\over \alpha} + \Delta(\alpha)}} \ee
where
\be \Delta(\alpha) = \int_{\alpha}^{\infty} {v\over { v-\alpha}} P(v) dv \ee
The parameter $ \alpha > 0$
when eliminated
between the above expressions gives $ J $ as a function of $ \rho $. We will
show that for a large class of distributions, $ J(\rho)$ is a convex function.
The only condition that $ P(v)$ should satisfy is that $
\lim_{v\longrightarrow \alpha} {P(v) \over {(v-\alpha)^2}}= 0 $.
To prove this we rewrite (20) as:
\be J =\alpha( 1-\rho)\hskip 1cm {\rho \over (1-\rho)} =
 \alpha \int_{\alpha}^{\infty} {v\over { v-\alpha}} P(v) dv \ee
from which $ {d^2J\over {d \rho^2}} = {d^2{\alpha}\over {d \rho^2}}(1-\rho)
- 2 {d \alpha\over {d \rho}} $.
Denoting the integral $ \int_{\alpha}^{\infty} {v\over {( v-\alpha)^k}}
P(v) dv $  by $ I_k$; noting that $ {d\over d{\alpha}}I_k = k I_{k+1}\ \
 {\rm for}\ \  k=1,2$
and differentiating the second relation of (22) twice with respect to $ \rho $,
we find after some algebra
\be {d^2J\over d{\rho}^2} = -2 {1\over {(1-\rho)^3}}{I_2 + \alpha I_3 \over
{( I_1 + \alpha I_2)^2}}\ee
which is always negative. This then hints and only hints at the possiblity of shock
wave solutions
for the equaion $ { \partial \rho \over {\partial t}} + {\partial J \over
{\partial x}} = { \partial \rho \over {\partial t}}+ J'(\rho){ \partial \rho
\over {\partial x}}=0$, the simplest of which are in the form of step functions
with density $ \rho_-$ at the left and $ \rho_+ $ at the right of
the front which
itself moves with the speed $ V_{shock} = { J_- - J_+\over { \rho_- - \rho_+}}$.
In conclusion, the approach presented in this letter, that is, begining
from consistent  algebras
and then finding the process may be further pursued to study other one
 dimensional reaction diffusion processes.
It may also be possible to map this model to more complex interface
growth models.
In this letter we have been lucky to postulate an algebra which while
being simple and beautiful,approximately models traffic, an every-day
life phenomenon.

\section{Acknowledgement}
I have benefited very much from discussions with J. Davoodi whom
I thank hereby.
\newpage
{\large {\bf References}}
\begin{enumerate}

\item F. Spitzer, Adv. Math. {\bf 5},246(1970)
\item T. M. Ligget, Interacting Particle Systems (Springer-Verlag, New York, 1985)
\item H. Spohn, Large Scale Dynamics of Interacting Particles
(Springer-Verlag, New York, 1991)
Phys. Rev. A{\bf 34},5091
\item D. Dahr, Phase Transiton{\bf 9},51 (1987)
\item B. Derrida,Phys. Rep. {\bf 301}, 65 (1998)
\item T. Halpen-Healy and Y. C. Zhang, Phys. Rep.{\bf 254},215(1994)
\item M. Kardar, G. Parisi, and Y. C. Zhang,Phys. Rev. Lett. {\bf 56},889(1986)
\item J. Krug and H. Spohn in{\it Solids Far From Equilibrium}, C. Godreche,
ed. (Cambridge University Press,1991)
\item C. T. MacDonald, J.H.Gibbs, and A.C.Pipkin, Biopolymers {\bf 6},1(1968)
;C.T.MacDonald, J.H.Gibbs, Biopolymers {\bf 7},707(1969)
\item O. Biham, A.A.Middelton and D. Levine, Phys. Rev. A{\bf 46},6124(1992)
\item K. Nagel, Phys.Rev. E{\bf 53},4655(1996)
\item K. Nagel and M. Schreckenberg, J. Physique I{\bf @},2221(1992)
\item B. Derrida and M. R. Evans in {\it " Non-Equilibrium
Statistical Mechanics in
one Dimension}", V. Privman ed. (Cambridge University Press, 1997)
\item P. Meakin, P. Ramanlal, L.M. Sander and R. C. Ball,
\item B. Derrida and K. Mallik, J.Phys.A: Math. Gen.{\bf 30},1031(1997)
\item B. Derrida and M. R. Evans, J.Phys. I France {\bf 3},311(1993)
\item B. Derrida, E. Domany and D. Mukammel; J. Stat. Phys.{\bf 69} 667(1992)
\item G. Schutz and E. Domany; J. Stat. Phys.{\bf 72} 277(1993)
\item B. Derrida, M.R. Evans, V.Hakim and V. Pasquier, J.Phys.A:Math.Gen.
{\bf 26}1493(1993)
\item V. Hakim, J. P. Nadal, J.Phys. A ; Math. Gen. {\bf 16} L213 (1983)
\item R. B. Stinchcombe and G. M. Schutz; Phys. Rev. Lett.{\bf 75},140(1995);
Europhys. Lett. {\bf 29},663(1995)
\item Kreb K and Sandow S 1997 J.Phys. A ; Math. Gen. {\bf 30} 3165
Stochastic Systems and Quantum Spin Chains {\it preprint} Cond-mat/9610029
\item V. Karimipour, " A Multi-Species Asymmetric Simple Exclusion
Process and its
relation to Traffic Flow " preprint, cond-mat 9808220 ( to appear in Phys. Rev.
E)
\item E.Ben-Naim and P.L. Krapivsky; Maxwell Models of Traffic Flow
{\it preprint} Cond-mat/9808162;
Phys.Rev. E {\bf 56},6680 (1997)
\item Arndt P Heinzel T and Rittenberg V, 1998 J.Phys. A ; Math.
Gen. {\bf 31} 833
\item Alcaraz F C Dasmahapatra S and Rittenberg V, 1998 J.Phys.
 A ; Math. Gen. {\bf 31} 845

\end{enumerate}

\end{document}